\documentclass{elsart}
\usepackage{graphicx}
\usepackage{amsmath}
\usepackage{amssymb}
\usepackage{epsfig}
\usepackage{dcolumn}
\usepackage{bm}
\begin{document}
\begin{frontmatter}

\title{Statistical theory \break of self-similarly distributed fields}
\author{ Alexander Olemskoi}
\address{Institute of Applied Physics, Nat. Acad. Sci. of Ukraine \break 58, Petropavlovskaya St., 40030 Sumy,
Ukraine}
\author{Irina Shuda}
\address{Sumy State University \break 2, Rimskii-Korsakov St., 40007 Sumy, Ukraine}
\date{}

\begin{abstract}
A field theory is built for self-similar statistical systems with both
generating functional being the Mellin transform of the Tsallis exponential and
generator of the scale transformation that is reduced to the Jackson
derivative. With such a choice, the role of a fluctuating order parameter is
shown to play deformed logarithm of the amplitude of a hydrodynamic mode.
Within the harmonic approach, deformed partition function and moments of the
order parameter of lower powers are found. A set of equations for the
generating functional is obtained to take into account constraints and symmetry
of the statistical system.
\end{abstract}

\end{frontmatter}
\maketitle

\section{Introduction}\label{Sec.1}

A formal basis of the statistical theory, using quantum field methods, is known
to be a generating functional which presents the Fourier-Laplace transform of
the partition function from the dependence on the fluctuating distribution of
an order parameter to an auxiliary field \cite{Zinn}. Due to the exponential
character of this transform, determination of correlators of the order
parameter is provided by ordinary differentiation of the generating functional
over auxiliary field.

Above scheme becomes inconsistent with passage from simple systems to complex
ones because the phase space gets forbidden regions and the phase flow does not
ensure statistical mixing \cite{T}. As is known from the theory of critical
phenomena, analytical description of complex systems is achieved in the
presence of the scaling invariance only \cite{Sor}. Because in this case the
role of a basic function plays the power-law function instead of the
exponential one, we need in use of the Mellin transform at constructing of the
generating functional. Moreover, one should introduce the Jackson derivative as
a generator of the scaling transformation instead of the ordinary derivation
operator.

This Letter is devoted to building a field-theoretical scheme based on the use
of both Mellin transform and Jackson derivative. The work is organized as
follows. In Section \ref{Sec.2} we adduce necessary information from the theory
of quantum calculus. Section \ref{Sec.3} is devoted to construction of the
generating functional and finding its connection with related correlators. As
the simplest example, the harmonic approach is studied in Section \ref{Sec.4}
to obtain the partition function and the order parameter moments of the first
and second powers in dependence of the deformation parameter. In Section
\ref{Sec.5} we introduce pair of additive functional whose expansion into
deformed series yields both Green functions and proper vertices. Moreover, we
find here formal equations governing by the generating functional of systems
possessing a symmetry with respect to a field variation and being subjected to
an arbitrary constrain. Section \ref{Sec.6} concludes our consideration.

\section{Preliminaries}\label{Sec.2}

We begin by citing an information from the quantum calculus \cite{QG,QC} that
will be needed below. A basis of this calculus is the dilatation operator
$D_x^\lambda:=\lambda^{x\partial_x}$ being determined by the deformation
parameter $\lambda$ and the differentiation operator
$\partial_x\equiv\partial/\partial x$. Expanding formally the operator
$D_x^\lambda$ into the Taylor series, it is easily to define its action onto
the power-law function: $D_x^\lambda x^n=\left(\lambda x\right)^n$. Similarly,
the expansion of an analytical function $f(x)$ shows that, in correspondence
with the denomination, the operator $D_x^\lambda$ arrives at the dilatation
$\lambda$ of the argument of this function: $D_x^\lambda f(x)=f(\lambda x)$. A
set of eigen functions of the dilatation operator is reduced to a homogeneous
functions $h(x)$ defined by the equality $D_\lambda h(x)=\lambda^q h(x)$ with
the self-similarity degree $q$. In general case, this function takes the form
$h(x)=A_\lambda(x)x^q$ where a factor $A_\lambda(x)$ is obeyed the invariance
condition $A_\lambda(\lambda x)=A_\lambda(x)$. It is convenient to pass from
the dilatation operator to the Jackson derivative
\begin{equation}
\mathcal{D}_x^\lambda:=\frac{D_x^\lambda-1}{(\lambda-1)x}
 \label{1}
\end{equation}
in accordance with the commutation rule
$[\mathcal{D}_x^\lambda,x]=D_x^\lambda$. The action the Jackson derivative onto
the homogeneous function is given by the relations:
\begin{equation}
\big(x\mathcal{D}_x^\lambda\big)h(x)=[q]_\lambda h(x),\quad
[q]_\lambda\equiv\frac{\lambda^q-1}{\lambda-1}.
 \label{2}
\end{equation}
Basic deformed number $[q]_\lambda$ represents a generalization of the exponent
of the homogeneous function.

Principle peculiarity of self-similar statistical systems is that their
consideration is based on the use of the deformed logarithm and exponential
\cite{T}
\begin{equation}
\ln_q(x):=\frac{x^{1-q}-1}{1-q},\quad\exp_q(x):=
\left[1+(1-q)x\right]_+^{1\over 1-q}
 \label{3}
\end{equation}
where $[y]_+\equiv\max(0,y)$. Moreover, one needs to deform the sum and product
as follows:
\begin{equation} \label{4}
x\oplus_q y=x+y+(1-q)xy,\quad x\otimes_q
y=\left[x^{1-q}+y^{1-q}-1\right]_+^{1\over 1-q};
\end{equation}
here, in the last equality $x,y>0$. Respectively, the deformed product of $n>1$
identical multipliers gives the expression of the deformed power-law function:
\begin{equation} \label{5}
\underbrace{x\otimes_q x\otimes_q\dots\otimes_q
x}_n=\left[nx^{1-q}-(n-1)\right]_+^{1\over 1-q}.
\end{equation}
Making use of the rules (\ref{4}) shows that functions (\ref{3}) are obeyed the
conditions
\begin{equation}\label{6}
\begin{split}
\ln_q(x\otimes_q x)=\ln_q x+\ln_q y,&\quad\ln_q(xy)=\ln_q x\oplus_q\ln_q y;\\
\exp_q(x+y)=\exp_q(x)\otimes_q\exp_q(y),&\quad\exp_q(x\oplus_q
y)=\exp_q(x)\exp_q(y).
\end{split}
\end{equation}

\section{Generating functional}\label{Sec.3}

As mentioned above, the generating functional of self-similar systems is
defined by the Mellin transform
\begin{equation} \label{I}
\mathcal{Z}_q\{J\}:=\int \mathcal{Z}_q\{\phi\}\phi^{J-1}\{{\rm d}\phi\},\quad
\phi^{J-1}\{{\rm d}\phi\}\equiv\prod\limits_{i=1}^N\phi_i^{J_i-1}{\rm d}\phi_i
\end{equation}
where index $i$ runs over lattice sites with number $N\to\infty$.\footnote{To
escape complications related to continuum space \cite{Zinn} we use the lattice
model.} According to Ref. \cite{T}, the partition functional
$\mathcal{Z}_q\{\phi\}=\exp_q(-S\{\phi\})$ is reduced to the deformed
exponential with the exponent being inverse action $S=S\{\phi\}$ determined by
the order parameter distribution. Functional $\mathcal{Z}_q\{J\}$ can be
presented by the deformed series \cite{QC}
\begin{equation} \label{series}
\mathcal{Z}_{\lambda}\{J\}=\sum\limits_{n=0}^{\infty}\frac{1}{[n]_\lambda!}\sum_{i_1\dots
i_n} \mathcal{Z}^{(n)}_{i_1\dots
i_n}\left(J_{i_1}-\lambda^0\right)\left(J_{i_2}-\lambda\right)\dots
\left(J_{i_n}-\lambda^{n-1}\right)
\end{equation}
where basic deformed factorial
$[n]_\lambda!=[1]_\lambda[2]_\lambda\dots[n]_\lambda$ is determined by the
product of related numbers (\ref{2}), while the coefficients
\begin{equation}
\mathcal{Z}^{(n)}_{i_1\dots i_n}=\left.
\left(J_{i_1}\mathcal{D}_{J_{i_1}}^\lambda\right)\left(J_{i_2}\mathcal{D}_{J_{i_2}}^\lambda\right)
\dots\left(J_{i_n} \mathcal{D}_{J_{i_n}}^\lambda\right)
\mathcal{Z}_q\{J\}\right|_{J_{i_1},\dots,J_{i_n}=1}
 \label{K}
\end{equation}
are given by the Jackson derivative (\ref{1}). By definition,
\begin{equation} \label{D}
\left.\left(J_i\mathcal{D}_{J_i}^\lambda\right)
\phi^{J-1}\right|_{J_i=1}=\frac{\phi_i^\lambda-\phi_i}{(\lambda-1)\phi_i}\prod_{k\ne
i}\phi_k^{J_k-1}\equiv\sigma(\phi_i)\prod_{k\ne i}\phi_k^{J_k-1}
\end{equation}
where one denotes $\sigma(\phi_i)\equiv\ln_{2-\lambda}(\phi_i)$. Then, the
Jackson derivation of the functional (\ref{I})
\begin{equation} \label{FO}
\left.\left(J_i\mathcal{D}_{J_i}^\lambda\right)
\mathcal{Z}_q\{J\}\right|_{J_i=1}=\int\mathcal{Z}_q\{\phi\}\sigma_i(\phi_i){\rm
d}\phi_i\prod_{k\ne i}\phi_k^{J_k-1}{\rm d}\phi_k
=\int\sigma_i(\phi_i)\mathcal{Z}_q(\phi_i){\rm d}\phi_i,
\end{equation}
arriving at the function
$\mathcal{Z}_q(\phi_i):=\int\mathcal{Z}_q\{\phi\}\prod_{k\ne
i}\phi_k^{J_k-1}{\rm d}\phi_k$, yields the first moment of the deformed order
parameter:
\begin{equation} \label{OP}
\begin{split}
& \mathcal{S}\equiv\left<\sigma(\phi_i)\right>:=
Z_q^{-1}\int\ln_{2-\lambda}(\phi_i)\mathcal{Z}_q(\phi_i){\rm d}\phi_i,\\ &
Z_q\equiv\left.\mathcal{Z}_q\{J\}\right|_{\{J\}=1}=\int\mathcal{Z}_q\{\phi\}\{{\rm
d}\phi\}.
\end{split}
\end{equation}
It represents the Tsallis entropy at condition that a field $\phi_i$ is reduced
to the inverse probability to take a statistical state on a site $i$.
Respectively, the kernel (\ref{K}) determines the deformed correlator of an
arbitrary order $n$
\begin{equation}
\left<\sigma(\phi_1)\dots\sigma(\phi_n)\right>:=
Z_q^{-1}\int\mathcal{Z}_q(\phi_1,\dots,\phi_n)\prod_{m=1}^n\sigma(\phi_{i_m})
{\rm d}\phi_{i_m}=Z_q^{-1} \mathcal{Z}^{(n)}_{i_1\dots i_n}
 \label{KK}
\end{equation}
where the function
$\mathcal{Z}_q(\phi_1,\dots,\phi_n):=\int\mathcal{Z}_q\{\phi\}\prod_{k\ne
i_1,\dots,i_n}\phi_k^{J_k-1}{\rm d}\phi_k$ is used.

\section{Harmonic approach}\label{Sec.4}

Within the standard field-theoretical scheme \cite{Zinn}, the action $S=S_0+V$
is split into an unharmonic part $V=V\{\phi\}$ and the quadratic form
$S_0:=\frac{1}{2\Delta^2}\sum_i\phi_i^2$ determined by a variance $\Delta^2$.
In accordance with Eq. (\ref{6}), the partition functional is determined with
the equality $\exp_q(-S)=\exp_q(-S_0)\otimes_q\exp_q(-V)$ where expansion over
unharmonism with using the rule (\ref{D}) gives the expression
\begin{equation} \label{PT}
\mathcal{Z}_q\{J\}=\exp_q\left(-V\left\{J\mathcal{D}_J\right\}\right)
\otimes_q\mathcal{Z}_q^{(0)}\{J\}
\end{equation}
of the generating functional
\begin{equation} \label{I1}
\begin{split}
\mathcal{Z}_q\{J\}&:=\int\exp_q\left[-S\{\phi\}\right]\phi^{J-1}\{{\rm
d}\phi\}\\ &
=\int\exp_q\left[-S\{\phi\}\right]\exp\left\{J\ln(\phi)\right\}\{{\rm
d}\ln(\phi)\}
\end{split}
\end{equation}
in terms of the bare part $\mathcal{Z}_q^{(0)}\{J\}$ related to the action
$S_0$. The principle difference of the equality (\ref{PT}) from corresponding
expression for simple systems consists in deformation of both exponential
operator of perturbation and its action onto the bare functional. Explicit
expression of the generating functional (\ref{PT}) yields expansion into the
deformed series (\ref{series}) with the coefficients (\ref{K}) determining the
correlators (\ref{KK}).

Within framework of the harmonic approach, when the action
$S_0=\frac{1}{2\Delta^2}\sum_{i=1}^N\phi_i^2$ is reduced to the sum of the $N$
independent constituents, the functional $\mathcal{Z}_q^{(0)}\{J\}$ is
expressed throughout the site functions $z_q^{(0)}(J_i)$ by means of the
equality $\mathcal{Z}_q^{(0)}\{J\}=z_q^{(0)}(J_1)\otimes_q
z_q^{(0)}(J_2)\otimes_q\cdots\otimes_q z_q^{(0)}(J_N)$. Within the mean field
approach, all multipliers coincide so that the use of Eq. (\ref{5}) arrives at
the expression of the generating functional (\ref{I1}) in terms of the site
functions:
\begin{equation} \label{1G}
\ln_q\left[\mathcal{Z}_q^{(0)}\{J\}\right] \simeq
N\ln_q\left[z_q^{(0)}(J)\right].
\end{equation}
As expected, the deformed logarithm of the partition functional proves to be
additive value whose magnitude is determined by the cite constituent
\begin{equation} \label{G0}
z_q^{(0)}(J)=\frac{1}{2}\left(\frac{2\Delta^2}{1-q}\right)^{J/2} {\rm
B}\left(Q,\frac{J}{2}\right),\quad Q\equiv\frac{2-q}{1-q}.
\end{equation}
Here, the $\rm B$-function decays as $J^{-1}$ near the point $J=0$ transforming
into the power-law dependence $J^{-Q}$ in the limit $J\to\infty$. As a result,
the bare function (\ref{G0}) decays fast with the $J$ growing in vicinity of
the point $J=0$ and then increases exponentially. At $J=1$, the dependence
(\ref{G0}) gives the deformed partition function per one site
\begin{equation} \label{G1}
z_q^{(0)}=\sqrt{\Delta^2/2(1-q)}{\rm B}(Q,1/2).
\end{equation}
As a result, within the zero approach, the mean order parameter (\ref{OP})
takes the form
\begin{equation} \label{E1}
\mathcal{S}^{(0)}=-(\lambda-1)^{-1}
\left\{1-\left[\frac{\ln_q\left[z_q^{(0)}(\lambda)\right]}{\ln_q\left(z_q^{(0)}(1)\right)}
\right]_+^{\frac{1}{1-q}}\right\}.
\end{equation}

In the information theory, the principle role is played by the Fisher matrix
whose elements represent pair correlators of derivatives of the logarithm of
the probability distribution function with respect to parameters of this
function (in difference of the Tsallis entropy whose value gives a global
measure of uncertainty, the Fisher matrix determines a local measure of
information stored by the system) \cite{Inf}. In the absence of a space
correlations, such a measure is given by the moment of the second order
$\mathcal{C}\equiv\left<\left[\sigma(\phi_i)\right]^2\right>$ for which the use
of equations (\ref{KK}) and (\ref{K}) arrives at the expression
\begin{equation} \label{E2}
\mathcal{C}^{(0)}=(\lambda-1)^{-2}\left\{1-2\left[\frac{\ln_q\left[z_q^{(0)}
(\lambda)\right]}{\ln_q\left(z_q^{(0)}(1)\right)}
\right]_+^{\frac{1}{1-q}}+\left[\frac{\ln_q\left[z_q^{(0)}(\lambda^2)\right]}
{\ln_q\left(z_q^{(0)}(1)\right)} \right]_+^{\frac{1}{1-q}}\right\}.
\end{equation}

\begin{figure}[!h] \centering
\includegraphics[angle=0, width=110mm]{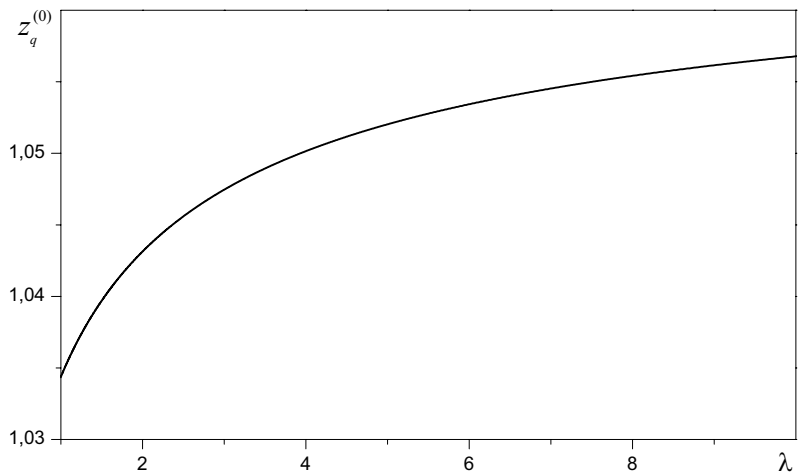}
\caption{Partition function (\ref{G1}) in dependence of the deformation
parameter.}\label{Z}
\end{figure}

\begin{figure}[!h]
\centering
\includegraphics[width=65mm]{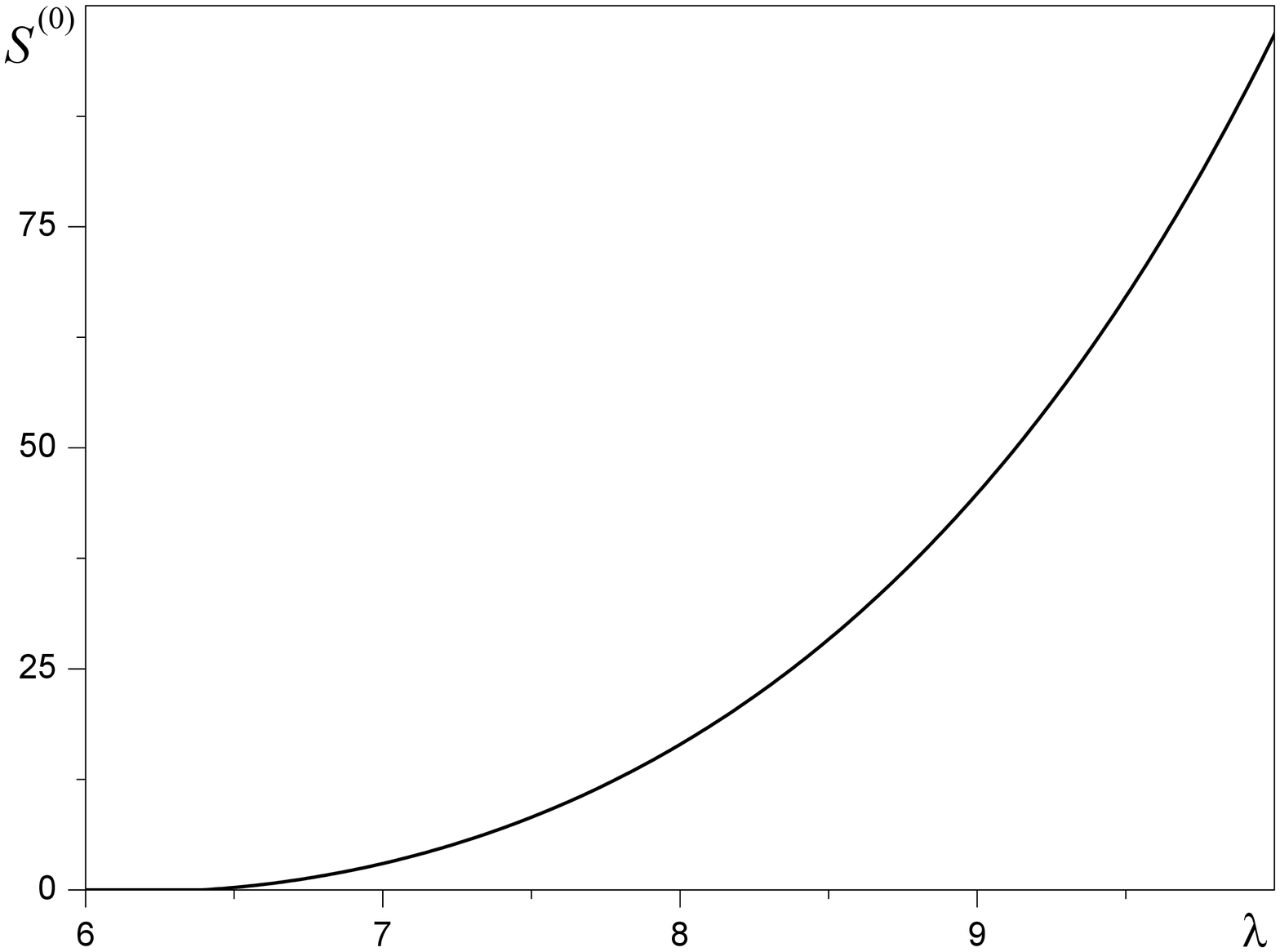}
\includegraphics[width=65mm]{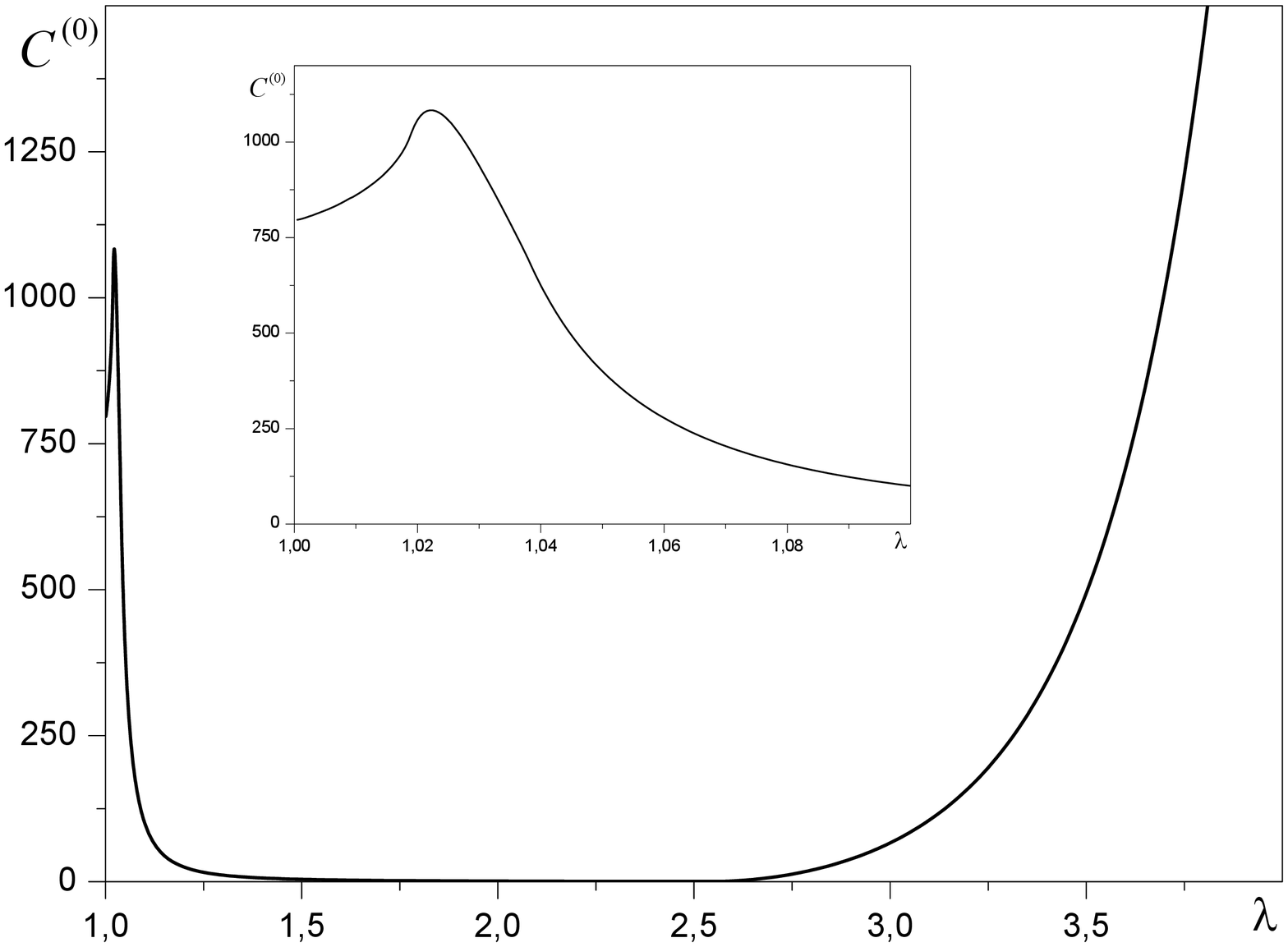}
\caption{Lower moments (\ref{E1}) and (\ref{E2}) of the deformed order
parameter.}\label{SC}
\end{figure}

At determination of the dependence of both moments (\ref{E1}) and (\ref{E2}) on
the deformation $\lambda$, one needs to take into account that parameter $q$ is
not free because a self-similarity condition restricts its value by the
equation \cite{q}
\begin{equation} \label{Qq}
\left(\lambda^{2-q}-1\right)\left(\lambda^{1-q}-1\right)=\left(\lambda-1\right)^2.
\end{equation}
As a result, variation of the deformation parameter $\lambda$ from $1$ to
$\infty$ arrives at growing the exponent $q$ since $0.382$ to $0.5$. According
to Figures \ref{Z} and \ref{SC}, hereby the specific
partition function $z_q^{(0)}\sim 1$ slightly increases, while the mean order
parameter (\ref{E1}) keeps the zero value before the deformation parameter
$\lambda=6.39$ and then increases monotonically. In contrast to this, the
second order moment (\ref{E2}) displays slight maximum in the region
$\lambda\geq 1$ that after downward excursion transforms into growing branch.
From the point of view of the information theory \cite{Inf}, this means that
before the value $\lambda=6.39$ a global measure of the information uncertainty
$\mathcal{S}$ does not appears, while the local measure of information stored
by the system $\mathcal{C}$ increases appreciably beginning since dilatations
$\lambda\sim 3$.

What about the correlators of the higher orders, in thermodynamic limit
$N\to\infty$, they are expressed in terms of the lower moments (\ref{E1}) and
(\ref{E2}) with help of different uncouplings. As a result, the use of
expression (\ref{PT}) allows one to build up a perturbation theory in analogy
with the standard scheme \cite{Zinn}.

\section{Field theory relations}\label{Sec.5}

If a system consists of macroscopically independent parts $1$ and $2$, then
related actions are connected with the additivity condition $S_{1+2}=S_1+S_2$,
whereas the deformed partition functional (\ref{I1}) whose kernel is determined
by the expression $\exp_q(-S_{1+2})=\exp_q(-S_1)\otimes_q\exp_q(-S_2)$ is
obeyed the deformed multiplicativity condition
$\mathcal{Z}_q^{1+2}=\mathcal{Z}_q^{1}\otimes_q\mathcal{Z}_q^{2}$. Therefore,
it is convenient to pass to the generating functional
\begin{equation}
\mathcal{G}_q:=\ln_q\left(\mathcal{Z}_q\right)
 \label{G}
\end{equation}
determined on the basis of the deformed logarithm that obeys the rules
(\ref{6}). As a result, the additivity condition
$\mathcal{G}_q^{1+2}=\mathcal{G}_q^{1}+\mathcal{G}_q^{2}$ becomes to be
satisfied so that the functional $\mathcal{G}_q=\mathcal{G}_q\{J\}$ can be
understood as a thermodynamic potential. Because the latter depends on an
auxiliary field $J$, one should use the Legendre transform
\begin{equation}
\Gamma_q\{\phi\}:=\sum_i J_i\sigma_i-\mathcal{G}_q\{J\},\quad
\sigma_i\equiv\ln_{2-q}(\phi_i)
 \label{L}
\end{equation}
to pass to a dependence on the order parameter $\phi$. This transform connects
conjugated pair of thermodynamic potentials $\mathcal{G}_q\{J\}$ and
$\Gamma_q\{\phi\}$ whose using arrives at the state equations
\begin{equation}
\sigma_i=\mathcal{D}_{J_i}^\lambda\mathcal{G}_q\quad\Leftrightarrow\quad
J_i=\mathcal {D}_{\sigma_i}^\lambda\Gamma_q.
 \label{SE}
\end{equation}
Similarly to the partition functional (\ref{I}), above potentials are presented
by the following deformed series:
\begin{equation}
\mathcal{G}_q\{J\}=\sum\limits_{n=1}^{\infty}\frac{1}{[n]_\lambda!}\sum_{i_1\dots
i_n}\mathcal{G}^{(n)}_{i_1\dots i_n}J_{i_1}\dots J _{i_n},
 \label{DJ}
\end{equation}
\begin{equation}
\Gamma_q\{\phi\}=\sum\limits_{n=1}^{\infty}\frac{1}{[n]_\lambda!}\sum_{i_1\dots
i_n} \Gamma^{(n)}_{i_1\dots i_n}\eta_{i_1}\dots\eta_{i_n},\quad
\eta_{i_m}\equiv\sigma_{i_m}-\lambda^{m-1}\mathcal{G}^{(1)}_{i_m} \label{DS}
\end{equation}
where in difference of the series (\ref{series}) terms with index $n=0$ are
absent (moreover, in Eq. (\ref{DJ}) expansion is carried out over the field
$J_{i_m}$ itself instead of the differences $J_{i_m}-\lambda^{m-1}$, while in
Eq. (\ref{DS}) appears the difference
$\eta_{i_m}\equiv\sigma_{i_m}-\lambda^{m-1}\mathcal{G}^{(1)}_{i_m}$,
$m=1,\dots,n$). With using the state equation (\ref{SE}), it is easily to
convince that coefficients of the series (\ref{DJ}) and (\ref{DS}) are
connected with the same relations which take place in simple systems
\cite{Zinn}. Within the diagrammatic representation, the kernels
$\mathcal{G}^{(n)}_{i_1\dots i_n}$ correspond to $n$-particle Green functions,
while $\Gamma^{(n)}_{i_1\dots i_n}$ determine proper vertices related
\cite{Zinn}.

Concluding this section, let us show that similarly to simple systems the
generating functional (\ref{I1}) obeys a set of formal equations. The first of
them is related to the system symmetry with respect to the field variation
$\delta\ln(\phi_i)=\epsilon f_i\{\phi\}$ being proportional to an arbitrary
functional $f_i\{\phi\}$ in the limit $\epsilon\to 0$. Due to this variation
the integrand of the functional (\ref{I1}) obtains the multiplier
$1+\epsilon\left[-\exp_q^q(-S){\partial
S}/{\partial\ln(\phi_i)}+J_i\right]f_i$, whereas the Jacobian of the passage
from the variable $\ln(\phi_i)$ to $\ln(\phi_i)+\delta\ln(\phi_i)$ equals
$1+\epsilon{\partial f_i}/{\partial\ln(\phi_i)}$ (the sum over repeated indices
is meant). Collecting all multipliers before the factor $\epsilon$, in
accordance with the invariance condition of the functional (\ref{I1}) one
finds:
\begin{equation} \label{EM2}
\begin{split}
\Bigg(f_i\left\{\frac{\delta}{\delta
J}\right\}\left[\exp_q^q\left(-S\left\{\frac{\delta}{\delta
J}\right\}\right)\frac{\partial
S}{\partial\ln(\phi_i)}\left\{\frac{\delta}{\delta J}\right\}-J_i\right]
\Bigg.\\ \Bigg.
 -\frac{\partial
f_i}{\partial\ln(\phi_i)}\left\{\frac{\delta}{\delta
J}\right\}\Bigg)\mathcal{Z}_q\{J\}=0.
\end{split}
\end{equation}
At $f_i\{\phi\}={\rm const}$, this equation is simplified to take the form
following immediately from the functional (\ref{I1}) after variation over the
integration variable $\ln(\phi)$.

The second of the pointed equations allows one to take into account an
arbitrary condition $F_i\{\ln(\phi)\}=0$ imposed on fields to be found.
Accounting this condition is achieved by introducing the $\delta$-functional
$\delta\{F\}$ into integrand of the last expression (\ref{I1}). As a result,
the generating functional takes the elongated form
\begin{equation} \label{I2}
\mathcal{Z}_q^{(F)}\{J\}:=\int
\exp_q\left[-S\{\phi\}\right]\exp\left\{J\ln(\phi)+\lambda
F\{\ln(\phi)\}\right\}\{{\rm d}\lambda\}\{{\rm d}\ln(\phi)\}.
\end{equation}
Variation of this expression over an auxiliary field $\lambda_i$ arrives at the
desired equation
\begin{equation} \label{EM3}
F_i\left\{\frac{\delta}{\delta J}\right\}\mathcal{Z}_q^{(F)}\{J\}=0.
\end{equation}

\section{Concluding remarks}\label{Sec.6}

Following the standard scheme \cite{Zinn}, we have considered the field theory
of self-similar statistical systems whose states are distributed in accordance
with the Tsallis exponential law. Because this distribution is characterized by
the power-law tail, we have used the generating functional (\ref{I}) based on
the Mellin transform and the Jackson derivative (\ref{1}) as the generator of
the scaling transformation. Along this line, the role of the order parameter
plays the mean value (\ref{OP}) of the deformed logarithm of the amplitude of a
hydrodynamic mode. The use of the harmonic approach shows the specific
partition function (\ref{G1}) slightly increases, the mean order parameter
(\ref{E1}) keeps the zero value before the deformation $\lambda\geq 6$ and then
increases monotonically, while the second order moment (\ref{E2}) displays
slight maximum in the region $\lambda\geq 1$ and after downward excursion
transforms into growing branch. Apart from the generating functional (\ref{I}),
we have introduced pair of the additive functional (\ref{G}) and (\ref{L})
whose expansions into deformed series (\ref{DJ}) and (\ref{DS}) yield both
Green functions and proper vertices. To take into account constraints and
symmetry of the statistical system we have obtained the equations (\ref{EM2})
and (\ref{EM3}) for the generating functional.

The special peculiarity of our consideration is that the kernel of the Mellin
transform (\ref{I}) is reduced to the Tsallis exponential (\ref{3}). But it is
worthwhile to stress that the Tsallis deformation is not sole of possible
procedures to obtain a distribution with power-law behavior. Another
possibility is known to be given by the basic deformed distribution \cite{BD}
that is invariant with respect to action of the Jackson derivative.\footnote{In
other words, eigen value of the Jackson derivative operator defined on the
basic deformed distribution equals $1$.} Moreover, generalized three-parameter
deformation procedure have been elaborated recently by Kaniadakis to obtain the
whole set of power-law tailed distributions \cite{K}. Building of generalized
field theory based on above pointed distributions is in progress.

\end{document}